\documentclass[3p,times,procedia]{elsarticle}
\usepackage{nupha_ecrc}

\usepackage{graphicx, subfigure, color}
\usepackage{dcolumn}
\usepackage{bm}
\usepackage{amsmath}

\def \anom {\rm{A}}

\newcommand {\beq}{\begin{eqnarray}}
\newcommand {\eeq}{\end{eqnarray}}

\volume{00}

\firstpage{1}

\journalname{Nuclear Physics A}

\runauth{Y. Hirono, D. Kharzeev, Y. Yin}

\jid{nupha}

\jnltitlelogo{Nuclear Physics A}

\usepackage{amssymb}

\usepackage[figuresright]{rotating}

\begin{document}

\begin{frontmatter}



\dochead{}

\title{New quantum effects in relativistic magnetohydrodynamics}


\author{Yuji~Hirono}

\address{Department of Physics, Brookhaven National Laboratory,
Upton, New York 11973-5000}

\author{Dmitri~E.~Kharzeev}
\address{Department of Physics and Astronomy, Stony Brook University, Stony Brook,
 New York 11794-3800}
\address{Department of Physics, Brookhaven National Laboratory, Upton, New York 11973-5000}
\address{RIKEN-BNL Research Center, Brookhaven National Laboratory, Upton, New York 11973-5000}
\author{Yi Yin}
\address{Center for Theoretical Physics, Massachusetts Institute of Technology, Cambridge 
MA 02139}

\begin{abstract}
Chiral anomaly induces a new kind of macroscopic quantum behavior in relativistic magnetohydrodynamics, including the chiral magnetic effect. In this talk we present two new quantum effects present in fluids that contain charged chiral fermions:
1) the turbulent inverse cascade driven by the chiral anomaly; 2) quantized chiral magnetic current induced by the reconnections of magnetic flux. We also discuss the implications for the evolution of the quark-gluon plasma produced in heavy ion collisions.
\end{abstract}


\end{frontmatter}
\vskip0.5cm

In this talk, we provide an elementary description of the recent papers \cite{Hirono:2015rla} and  \cite{Hirono:2016jps} on the quantum effects induced by the chiral anomaly in relativistic magnetohydrodynamics. Due to the lack of space, we refer the reader to these papers for details and a complete list of references.
\vskip0.1cm

The anomaly-induced transport of charge in systems with chiral fermions has attracted a significant interest recently. This interest stems from the possibility to study a new kind of a macroscopic quantum behavior. 
While the macroscopic manifestations of quantum mechanics are well known (for example, superfluids, superconductors and cold atoms), so far they have been mostly limited to systems with broken symmetries characterized by a local order parameter, e.g. the density of Cooper pairs in superconductors. 
The effects induced by the chiral anomaly \cite{Adler:1969gk,Bell:1969ts}
in systems with chiral fermions are of different nature. 

Let us consider as an example the Chiral Magnetic Effect (CME) in systems with charged chiral fermions -- the generation of electric current in an external magnetic field
induced by the chirality imbalance \cite{Kharzeev:2004ey}, see Refs.~\cite{Kharzeev:2013ffa,Kharzeev:2015znc} for reviews and references. The experimental observation of CME in a Dirac semimetal ${\rm ZrTe}_5$ has been reported recently in  \cite{Li:2014bha}; currently, over a dozen of 3D chiral materials have been shown to exhibit the CME, see e.g. \cite{Xiong2015,Huang2015}.
In this case, no symmetry has to be broken, and the system is in its normal state. However the chirality imbalance is linked by the Atiyah-Singer theorem to the non-trivial global topology of the gauge field. Since the global topology of the gauge field cannot be determined by a local measurement, there is no corresponding local order parameter, and we deal with ``topological order".

This has very important implications for the real-time dynamics of a system composed by charged chiral fermions and a dynamical electromagnetic field. Let us initialize the system by creating a lump of chirality imbalance localized within a magnetic flux forming a closed loop, see Fig. 1a. The magnetic field will induce the CME current flowing along the lines of magnetic field $\bf{B}$ (an effect that is absent in Maxwell electromagnetism). Because the electric  CME current itself acts as a source for the magnetic field, the current flowing along $\bf{B}$ will twist the lines of magnetic flux (see Figs. 1b,c). This will induce a non-zero expectation value for the {\it magnetic helicity} known since Gauss's work in XIX century and introduced in magnetohydrodynamics by Woltjer \cite{Woltjer} and Moffatt \cite{Moffatt}, see also \cite{ArnoldKhesin}: 
\beq
\label{mag_hel}
h_m \equiv \int d^3 x\ \bf{A} \cdot \bf{B} \, . 
\eeq
Magnetic helicity is a topological invariant (Chern-Simons three-form) characterizing the global topology of the gauge field. It is mathematically related to the knot invariant, and measures the chirality of the knot formed by the lines of magnetic field. Because of this, the generation of magnetic helicity will create the chiral knot out of the closed loop of magnetic flux -- so the topology of magnetic flux will change. In a recent paper \cite{Hirono:2015rla} we have quantified this statement, and studied how the topology of magnetic flux changes in real time. We have found that as a consequence of chiral anomaly and the CME, the magnetic field evolves to the self-linked Chandrasekhar - Kendall states (see Fig. 1d), that are solutions of the ${\bf \nabla} \times {\bf B} \sim {\bf B}$ equation.
\begin{figure}[htbp]
\centering
\subfigure[]
{
   \label{fig:untwisted} 
        \includegraphics[width=3cm]{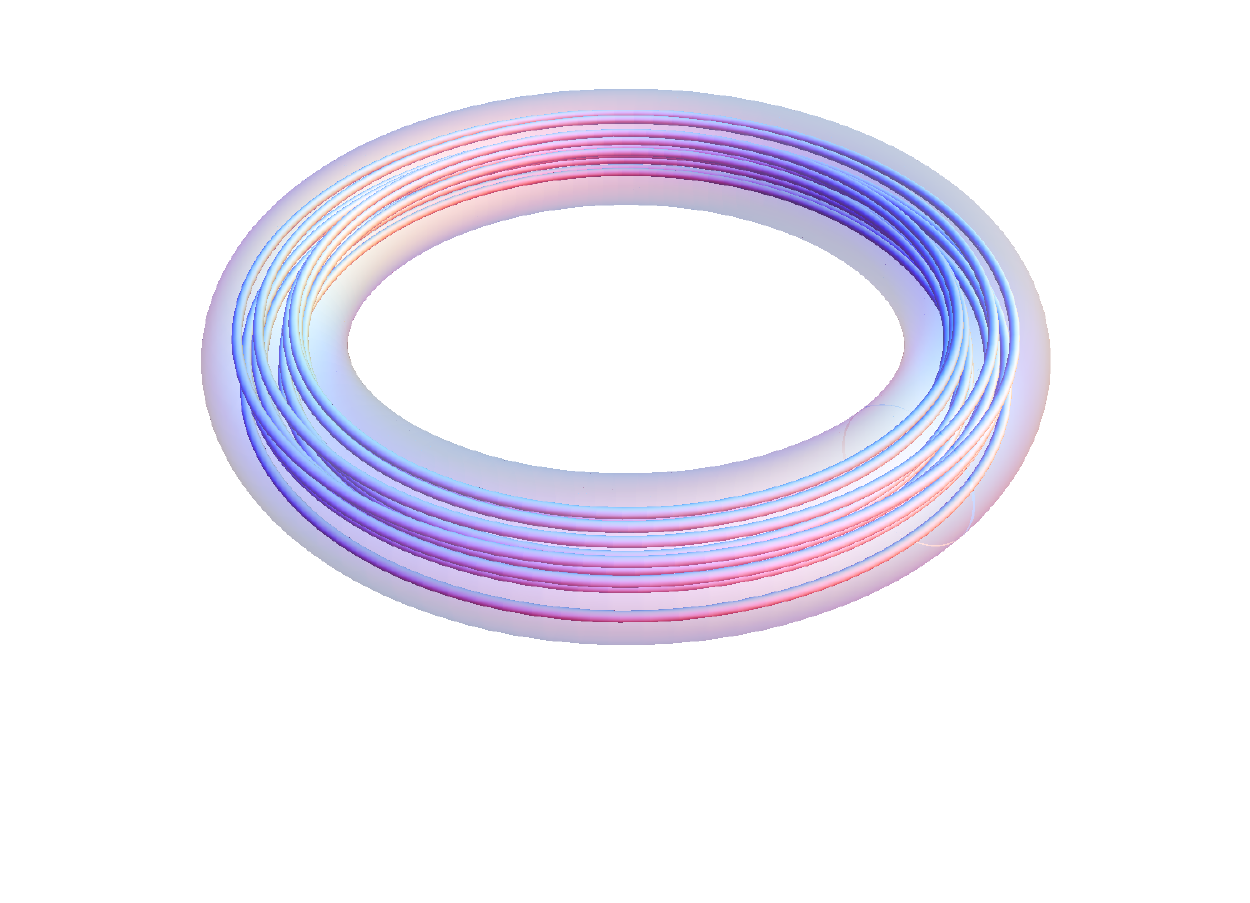}
} 
\subfigure[]
{
   \label{fig:twisted} 
        \includegraphics[width=3cm]{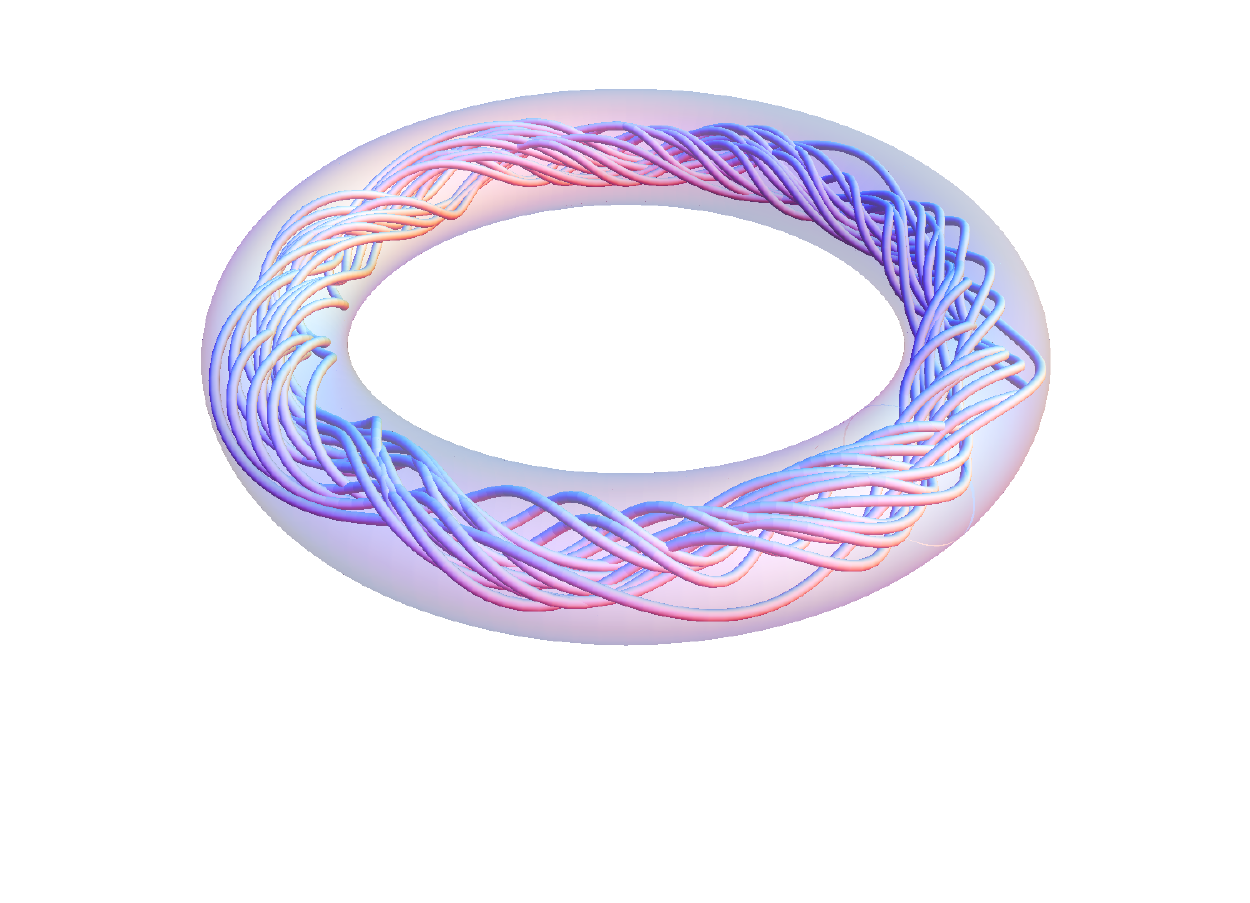}
} 
\subfigure[]
{
   \label{fig:trefoil} 
        \includegraphics[width=3cm]{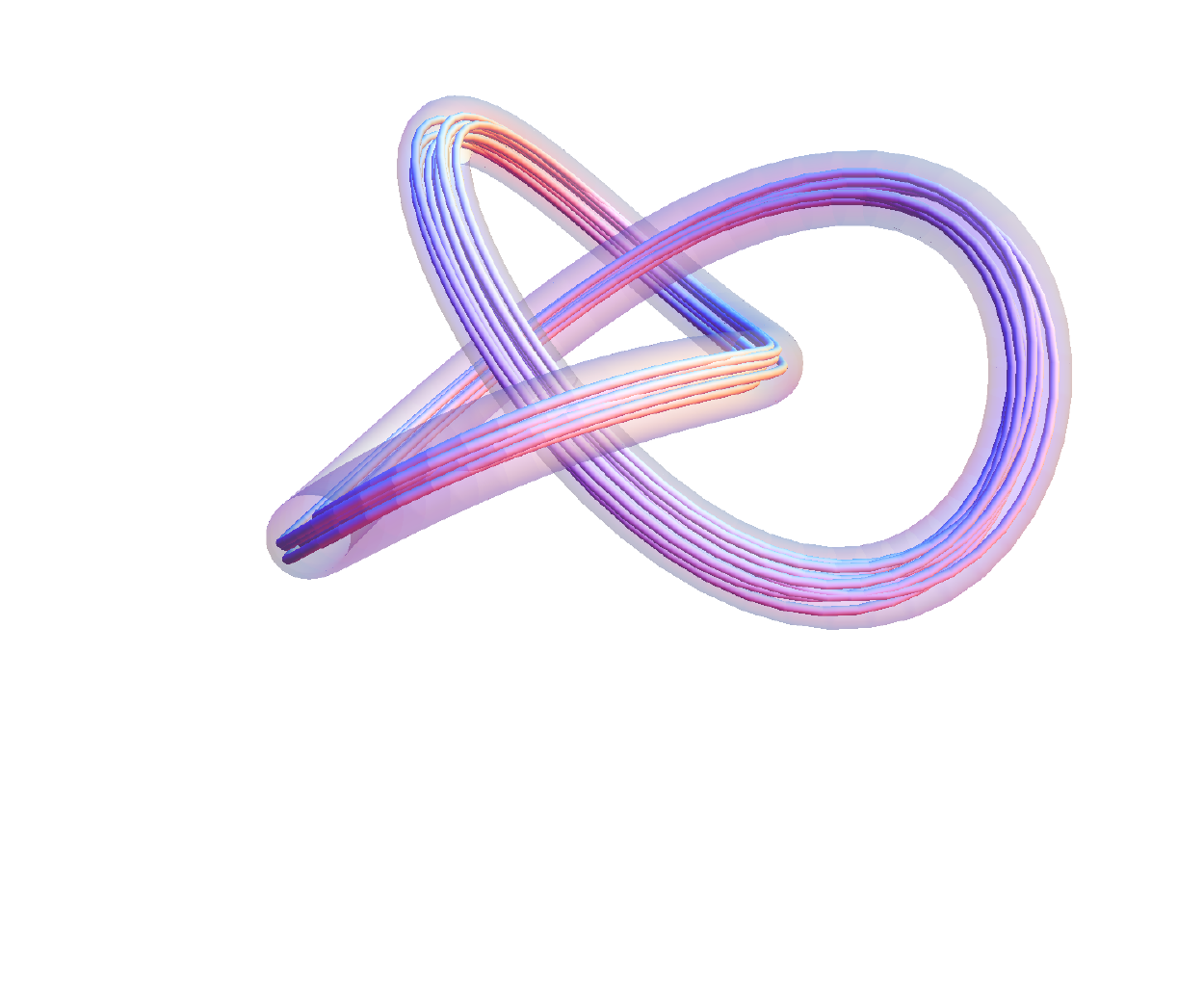}
} 
\subfigure[]
{
   \label{fig:CK} 
        \includegraphics[width=3cm]{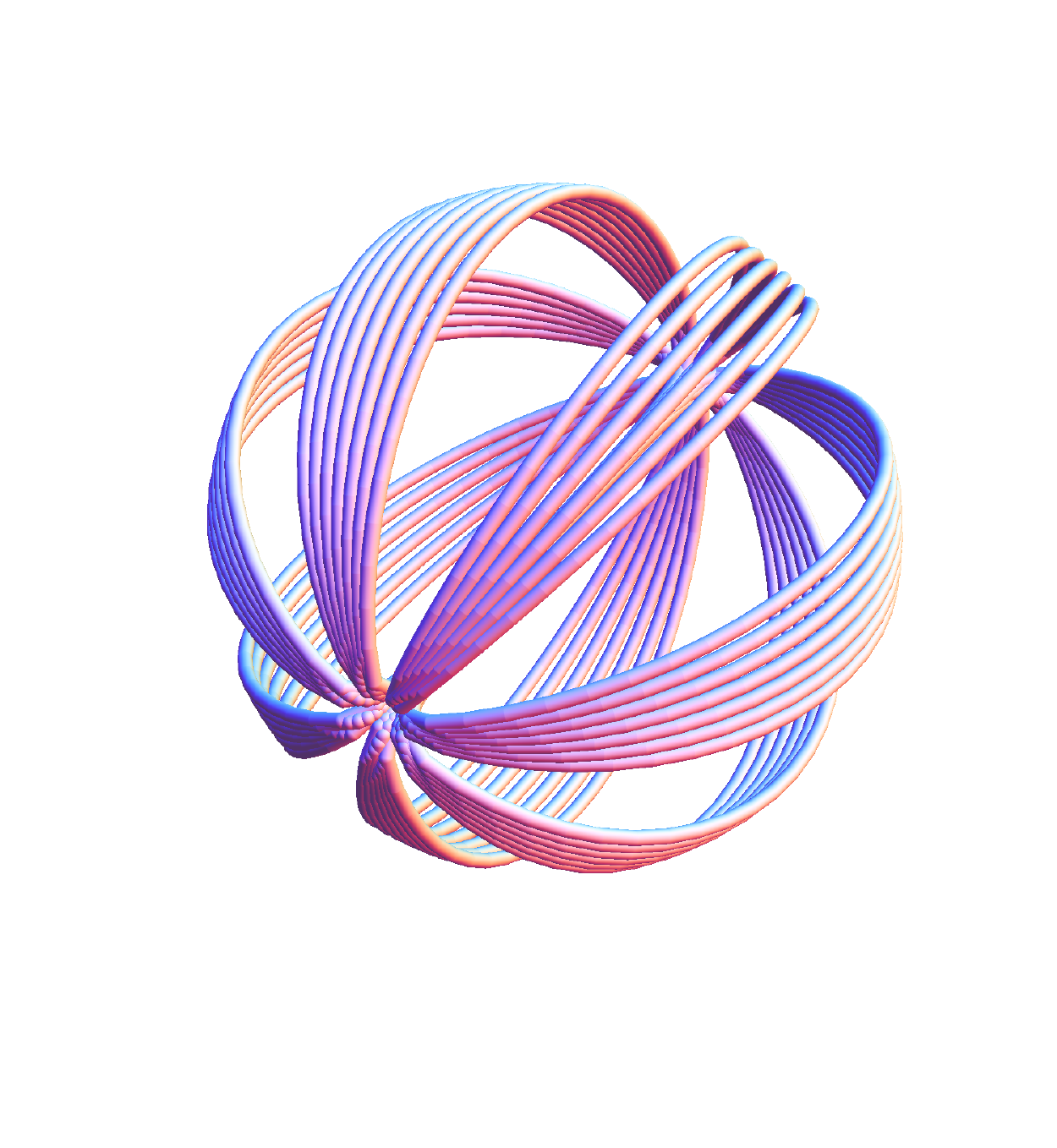}
} 
\caption{
\label{fig:topology}
The topology of Abelian magnetic flux: (a) upper left -- untwisted loop; 
(b) upper right -- twisted magnetic flux; 
(c) lower left -- the self-linked magnetic flux (trefoil knot shown); 
(d) lower right -- the self-linked  Chandrasekhar-Kendall state.
}
\end{figure}

During the evolution, the size of the knot of magnetic flux increases due to the chiral magnetic instability. Moreover, at late times this evolution becomes self-similar, and is characterized by universal exponents \cite{Hirono:2015rla}. 
To discuss the qualitative picture of the inverse cascade driven by anomaly, 
let us first define the {\it fermionic helicity}:
\beq
\label{hF_def}
h_{F}\equiv C^{-1}_{\anom}\int d^{3}x\,\ n_{A}\, ,
\eeq
where $n_A = j^0_A$ is the density of axial charge. 
The total helicity of the system $h_{0}$ is conserved: 
\beq
\label{h0}
h_{0}\equiv 
h_{m}+h_{F}
=\text{const}\, ,
\eeq
but the chiral anomaly can re-distribute helicity between the fermionic and magnetic parts.
Therefore, the 
system will tend to minimize the energy cost at a fixed helicity. This has been found to induce the inverse cascade of magnetic helicity \cite{Joyce:1997uy,Boyarsky:2011uy,Tashiro:2012mf,Tuchin:2014iua,Manuel:2015zpa} . As recently shown in \cite{Hirono:2015rla}, 
the inverse cascade is achieved by transferring helicity to the Chandrasekhar-Kendall states of increasing size. 
Moreover, the corresponding inverse cascade of magnetic helicity is self-similar, and is characterized by universal exponents that correspond to the diffusion universality class. The self-similarity of the cascade is illustrated in Fig.2: the inverse cascade of the momentum spectrum $g(k,t)$ of magnetic helicity $h_m$ defined by $h_m(t) = 1/\pi \int dk\ k\ g(k,t)$ shown in Fig.2a exhibits a self-similar behavior, see Fig. 2b. The values of the exponents are fixed to $\alpha = 1$, $\beta = 1/2$ by the dynamics of the inverse cascade and dimensional arguments, see \cite{Hirono:2015rla}.  This self-similar cascade exists also when the velocity of the fluid is taken into account, as shown in \cite{Yamamoto:2016xtu}.
\begin{figure}
\centering
       \subfigure[]{
        \includegraphics[width=0.45\textwidth]{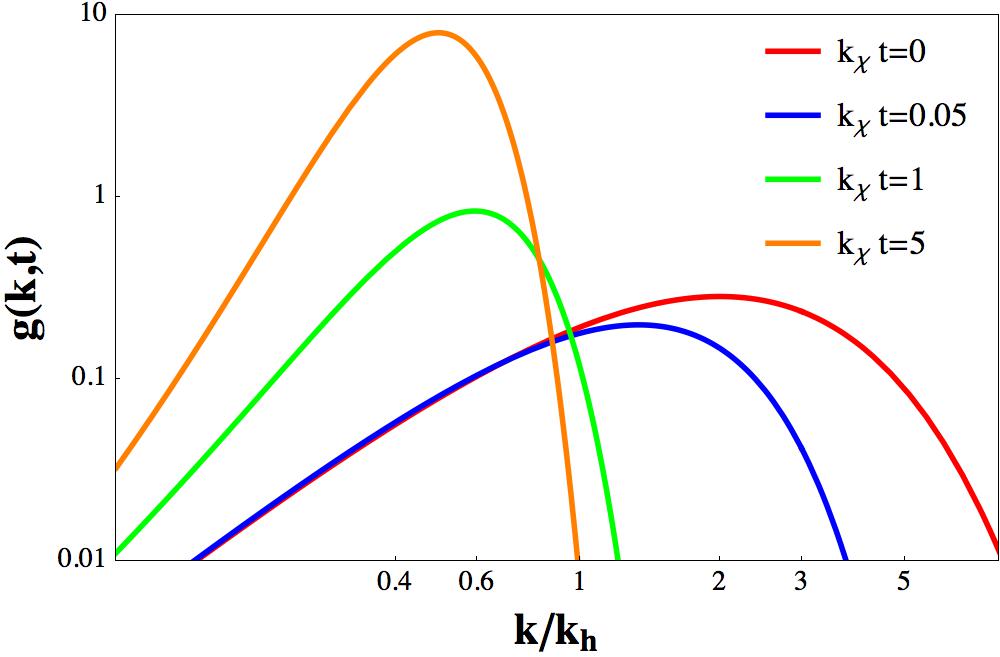}
         \label{fig:gstage}
      }
       \subfigure[]{
        \includegraphics[width=0.45\textwidth]{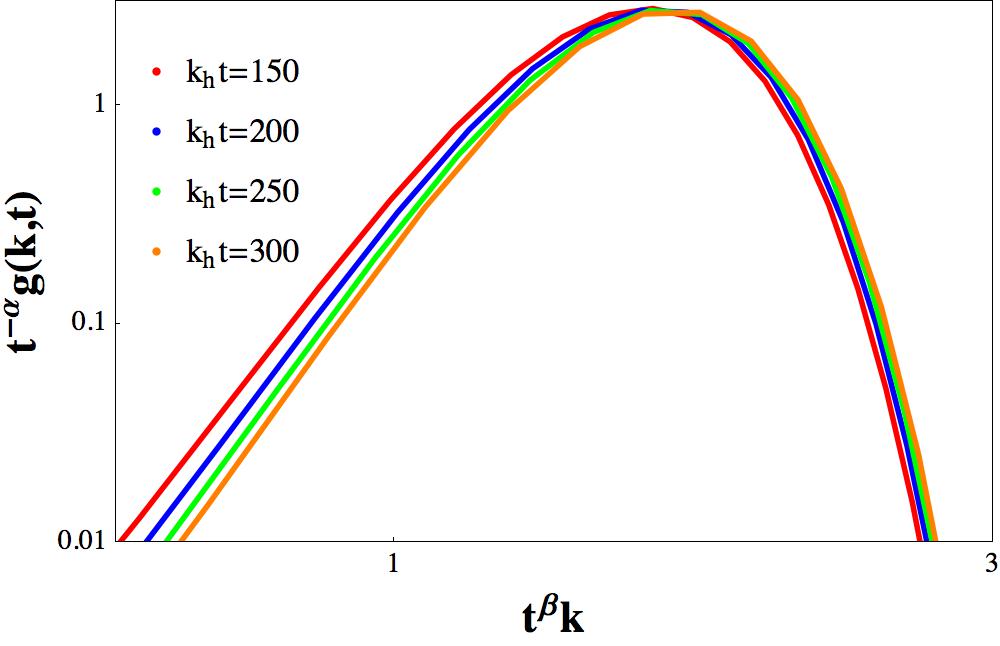}
                \label{fig:gsimilar}
      }
\caption{
\label{fig:bevol}
(Color online) The evolution of magnetic helicity spectrum $g(k,t)$.
(Left): $g(k,t)$ at initial time $t=0$ (red) and three representative
times respectively.
(Right): $t^{-\alpha}g(k,t)$ vs $t^{\beta}k$ in the self-similar stage of the
evolution. 
}
\end{figure}

\begin{figure}[htbp]
\centering
\includegraphics[width=0.5\textwidth]{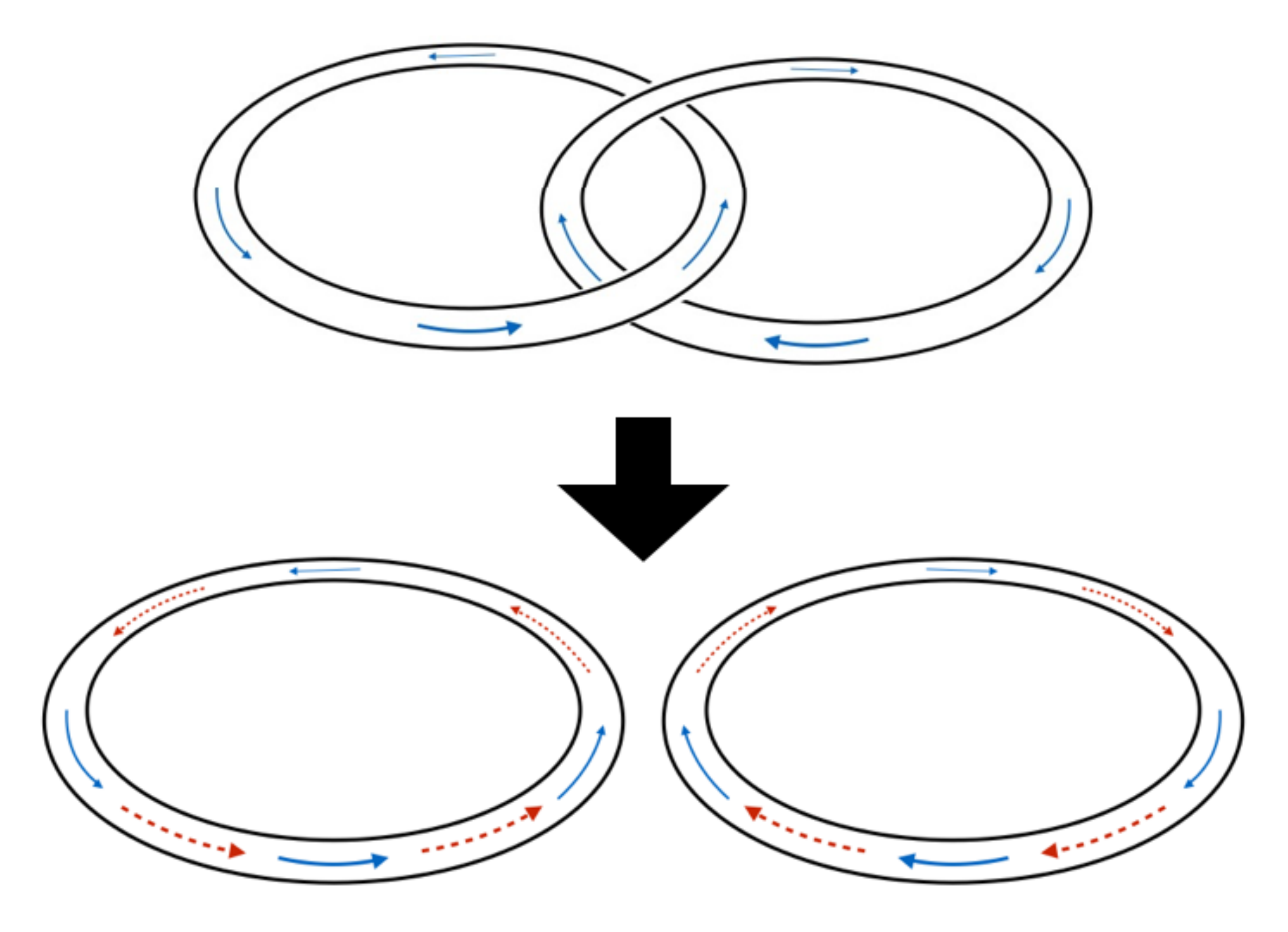}
\caption{
\label{fig:unlink1}
 (Color online)
 Current generation associated with unlinking of a simple link of two
 flux tubes.
 The solid arrows denote the directions of the magnetic field,
 and the dotted arrows indicate the directions of generated CME currents. 
 }
\end{figure}

So far, we relied on an external source of fermionic helicity to initiate the inverse cascade. An initial fermionic helicity can be provided by coupling the theory to a non-Abelian sector, where the instanton and sphaleron solutions exist. 
This corresponds to the conventional scenario in heavy ion physics where sphalerons in hot QCD plasma create a chirality imbalance that then induces the chiral magnetic effect in an external abelian magnetic field created by the colliding ions \cite{Kharzeev:2004ey}. However, as shown in \cite{Hirono:2016jps}, the chirality imbalance can be created by the topology of abelian magnetic field within resistive relativistic magnetohydrodynamics.

Indeed, consider the 
chirality associated with the topology of magnetic flux. In the absence of magnetic monopoles the lines of
magnetic field are closed. For example, the field lines of a
solenoid form an ``unknot''. However the topology of magnetic flux can be
more complicated, and magnetic flux can form a chiral
knot. Magnetic reconnections can change the chirality of this knot, and thus 
induce the imbalance of chirality in the system, see Fig.3. This imbalance of
chirality will then lead to the generation of the chiral magnetic current. The corresponding
chiral magnetic current is quantized, and is completely determined
by the knot invariants \cite{Hirono:2016jps}.

The main result of \cite{Hirono:2016jps} is the following formula for the generated current $\Delta \bm J$ 
along the loops $C_i$ of magnetic flux  in terms of the change 
$\Delta \mathcal H$ of the magnetic helicity,
which is a topological measure of the knot (to be defined below), 
\begin{equation}
\sum_{i} \oint_{C_i} \Delta \bm J \cdot d \bm x = - \frac{e^3}{2 \pi^2}
\Delta \mathcal H ,
\label{main_res}
\end{equation}
where $e$ is electric charge. Since $\Delta \mathcal H$ is
an integer number times the flux squared,
the CME current resulting from  reconnections of magnetic flux is
quantized.
The process illustrated in Fig. \ref{fig:unlink1} shows the simplest
realization of such currents. 
The unlinking of a link involves the topology change of the magnetic fluxes 
which generates the CME currents (indicated by dotted arrows) on both tubes.
The amount of the integrated current over the tubes is given by the helicity
change during the process, as quantified by Eq.~(\ref{main_res}).

 In heavy ion collisions, the reconnections of magnetic flux can thus provide a source for the chiral magnetic current during the late hydrodynamical stage of the quark-gluon plasma evolution. Since the electrical conductivity of the quark-gluon plasma is finite, the magnetic helicity is not conserved, and so the reconnections of magnetic flux should occur. Apart from the effect on charged hadron asymmetries, the chiral magnetic currents propagating along the curved lines of magnetic field will act as sources of soft photons. A quantitative study of these and related CME effects will require a numerical description combining dynamical electromagnetic fields \cite{Inghirami:2016iru} with the currents induced by the chiral anomaly. This work is part of the Beam Energy Scan Theory (BEST) Collaboration research plan, and is ongoing. 
\vskip0.1cm

This work was supported in by the U.S.
Department of Energy under Contracts No. DE-FG- 88ER40388, DE-AC02-98CH10886 and DE-SC0011090 and  within the framework of the Beam Energy Scan Theory (BEST) Topical Collaboration.
\bibliographystyle{elsarticle-num}







\end{document}